\documentclass{elsart}
\usepackage{graphicx}
\usepackage[figuresright]{rotating}

\input epsf

\begin{document}

\begin{frontmatter}

\title{Energy Calibration of the JLab Bremsstrahlung Tagging System}

\author{S. Stepanyan\thanksref{jlab}\corauthref{cor},} 
\author{S. Boyarinov\thanksref{jlab},}
\author{H. Egiyan\thanksref{jlab}\thanksref{unh},}
\author{L. Guo\thanksref{jlab},}
\author{D. Dale\thanksref{kentucky},}
\author{M. Gabrielyan\thanksref{kentucky},} 
\author{L. Gan\thanksref{unc},}
\author{A. Gasparian\thanksref{ncat},}
\author{A. Glamazdin\thanksref{khark},} 
\author{B. Mecking\thanksref{jlab},} 
\author{I. Nakagawa\thanksref{kentucky}\corauthref{cur},}
\author{A. Teymurazyan\thanksref{kentucky},}
\author{M.H. Wood\thanksref{umass}}

\corauth[cor]{Corresponding author. Address: Jefferson Lab, 12000
Jefferson Ave., Newport News, VA 23606. 
E-mail address: stepanyan@jlab.org}

\corauth[cur]{Current address: The Institute of Physical and Chemical 
Research (RIKEN), Wako, Saitama 351-0198, Japan}

\address[jlab]{Thomas Jefferson National Accelerator Facility,
Newport News, VA 23606, USA}

\address[unh]{University of New Hampshire, Department of Physics, 
Durham, NH 03824, USA}

\address[kentucky]{University of Kentucky, Department of Physics and
Astronomy, Lexington, KY 40506, USA}

\address[khark]{Kharkov Institute of Physics and Technology, Kharkov 61108, Ukraine}

\address[umass]{University of Massachusetts, Department of Physics, Amherst, 
MA 01003, USA}

\address[unc]{University of North Carolina at Wilmington, S. College
Rd., Wilmington, NC, 28403}

\address[ncat]{North Carolina A\&T State University, Greensboro, NC 27411}

\date{\today}

\begin{abstract}
\noindent
In this report, we present the energy calibration of the Hall B
bremsstrahlung tagging system at the Thomas Jefferson National
Accelerator Facility. 
The calibration was performed using a magnetic pair spectrometer. The tagged
photon energy spectrum was measured in coincidence with $e^+e^-$
pairs as a function of the pair spectrometer magnetic
field. Taking advantage of the internal linearity of the pair spectrometer,
the energy of the tagging system was calibrated at the level of $\pm
0.1\% E_\gamma$. The absolute energy scale was determined using the
$e^+e^-$ rate measurements close to the end-point of the photon
spectrum. The energy variations across the full tagging range 
were found to be $<3$ MeV.
\end{abstract}

\begin{small}
\leftline{} 
\end{small}

\end{frontmatter}

\newpage
\tableofcontents
%\maketitle

\section{Introduction}

\noindent

In this report, we present the method and the results of the energy
calibration of the Hall B 
photon tagging system \cite{tagger} at the Thomas Jefferson National
Accelerator Facility (JLab). The Hall B tagging system provides tagged
photons in the multi-GeV energy range with high energy resolution
($\sim 0.1\%E_\gamma$) and a broad tagging range, $20\%$ to $95\%$ of
$E_0$. It is used for the investigation of real 
photon induced reactions, primarily in conjunction with the CEBAF Large
Acceptance Spectrometer (CLAS) \cite{clas}. The tagging range of the device is
divided into $767$ energy bins ($E$-bins). In each $E$-bin, the
central value of the energy is used as the energy of the radiated photon. 
These values were generated by a ray-tracing program using the design 
geometry of the scintillation counter hodoscope ($384$ overlapping counters,
called $E$-counters) and the two-dimensional field map of the dipole magnet.

In the analysis of fully exclusive reactions, such as photoproduction on
deuterium\footnote{In the kinematically complete reaction $\gamma d\to
p\pi^+\pi^- (n)$ the neutron mass was used as a constraint to 
calculate the photon energy from the momenta of the charged particles
\cite{s1}.}, it was found that the photon 
energy defined by CLAS  and the central values of the $E$-bins of
tagged photons are different by as much as $0.5\%$.  
The variation of this difference as a function of the tagger
$E$-counter position was compatible with
a possible sag of the frames which hold the 
$E$-counters. There was corroborating evidence from simulations of the
effects of gravitational sagging and various possible misalignments of the
tagger focal plane \cite{taggeom}.

To determine corrections to the central value of energy in each $E$-bin 
independent of CLAS, the
tagged photon energy spectrum was measured in coincidence with
$e^+e^-$ pairs detected in the Hall-B pair 
spectrometer (PS)\cite{primex}. The data were collected as a function of
the PS dipole field and at fixed geometry of the $e^+e^-$ detectors. These
measurements, taking advantage of the linear relationship between
momentum and the magnetic field of the PS, allowed 
for the relative calibration of $E$-bins with high accuracy. Using  
measurements of $e^+e^-$ coincidence rates at PS settings close to the
end-point energy, 
the absolute energy scale of the tagging system was calibrated as well.
The energy correction factor for each $E$-bin is defined as the ratio
of photon energy, determined from momenta of $e^+e^-$ pairs, to the
ideal central value of the energy of that $E$-bin.

To calculate the momenta of the $e^+$ and $e^-$, a simplified model
for a homogeneous dipole was employed, using the central value of the field and the
positions of the particle trajectories at the entrance and exit
of the field region. Two corrections were introduced to the model 
based on ray-tracing simulations using measured and generated field
maps to account for non-linearities in the field
distribution and for the finite beam and detector sizes. 

For these measurements, the PS was instrumented with micro-strip detectors for
better position determination of $e^+$ and $e^-$, and thus better
energy resolution. 
Based on the level of knowledge of the PS dipole field
distribution and the position of the micro-strip detectors, 
the accuracy of the method is estimated to be $\delta E_\gamma\sim\pm 0.1\%E_\gamma$.

\section{Experimental setup and the measurements}

\noindent

\begin{sidewaysfigure}[htb]
\vspace{70mm} 
{\includegraphics{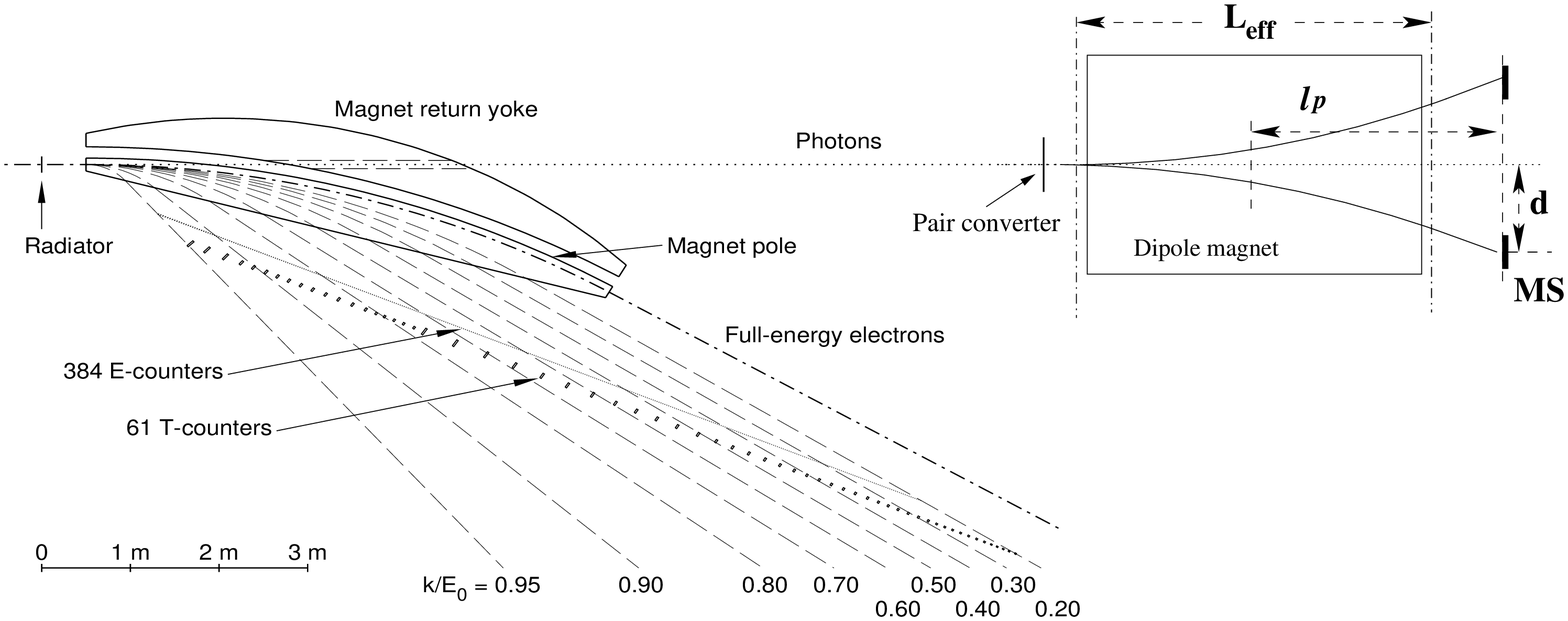}}
\caption{Schematic view of the setup. The focal plane
hodoscope consists of two planes of scintillation counters. The first
plane, called E-plane, contains $384$
overlapping counters, E-counters, and defines the energy bins. The second
plane, called T-plane, contains
$61$ counters, T-counters, used for the time coincidence with
CLAS. The pair production converter of the pair spectrometer is located about $8$ 
meters downstream of the radiator. For illustration purposes the
diagram of the pair spectrometer is rotated by $90^o$ around the beam axis.}
\label{fig:model}
\end{sidewaysfigure}

The data were collected during a photoproduction experiment in April of
2004 using the Hall B bremsstrahlung tagging system and the pair
spectrometer.
The description of the tagging system can be
found in Ref. \cite{tagger}, a schematic view of
the setup is shown in Fig. \ref{fig:model}. The photon beam was
generated in the interaction of a $E_e=3.776$ GeV electron beam with
a $10^{-4}$ radiation length thick $Au$ foil (``Radiator''). The tagger dipole magnet was
operated at $1273$ A 
current (corresponding to the central field value of $1.0627$ T) and
covered the tagged photon energy range 
from $0.9$ to $3.6$ GeV. The Hall B pair spectrometer is located 
$\sim 10$ meters downstream of the ``Radiator.'' The PS consists of a dipole
magnet and two planes of scintillation counters,
positioned symmetrically to the left and the right of the beam axis in
the horizontal plane downstream of the magnet. The pair production
converter, ``Pair converter'', is located 
$55.8$ cm upstream of the PS dipole center and consists of
$10^{-3}$ radiation length thick Aluminum.

For this measurement, in addition to the scintillation detectors, the PS
was instrumented with two pairs of micro-strip detectors (MS) to
provide better position determination for the $e^+e^-$ behind the
dipole. They were mounted $93.07$ cm downstream of the magnet
center, in front of the scintillation counters. Each pair of
micro-strip detectors consisted of $X$ and $Y$ planes and covered $20\times
20$mm$^2$ of detection acceptance. (The ($XZ$) plane is defined by the
centerline of the beam and the deflection plane of the dipole. The
main field component is parallel to the $Y$ axis).
The distance between the centroids of the ``X'' planes was $450\pm 0.5$mm,
and they were centered on the beam axis. The pitch size 
of the micro-strip detectors was $50\mu m$. In the off-line analysis, only
one of the ``Y'' planes was used, since the micro-strip for the second ``Y''
plane had a high noise level. 

\begin{figure}[htb]
\vspace{80mm} 
{\includegraphics{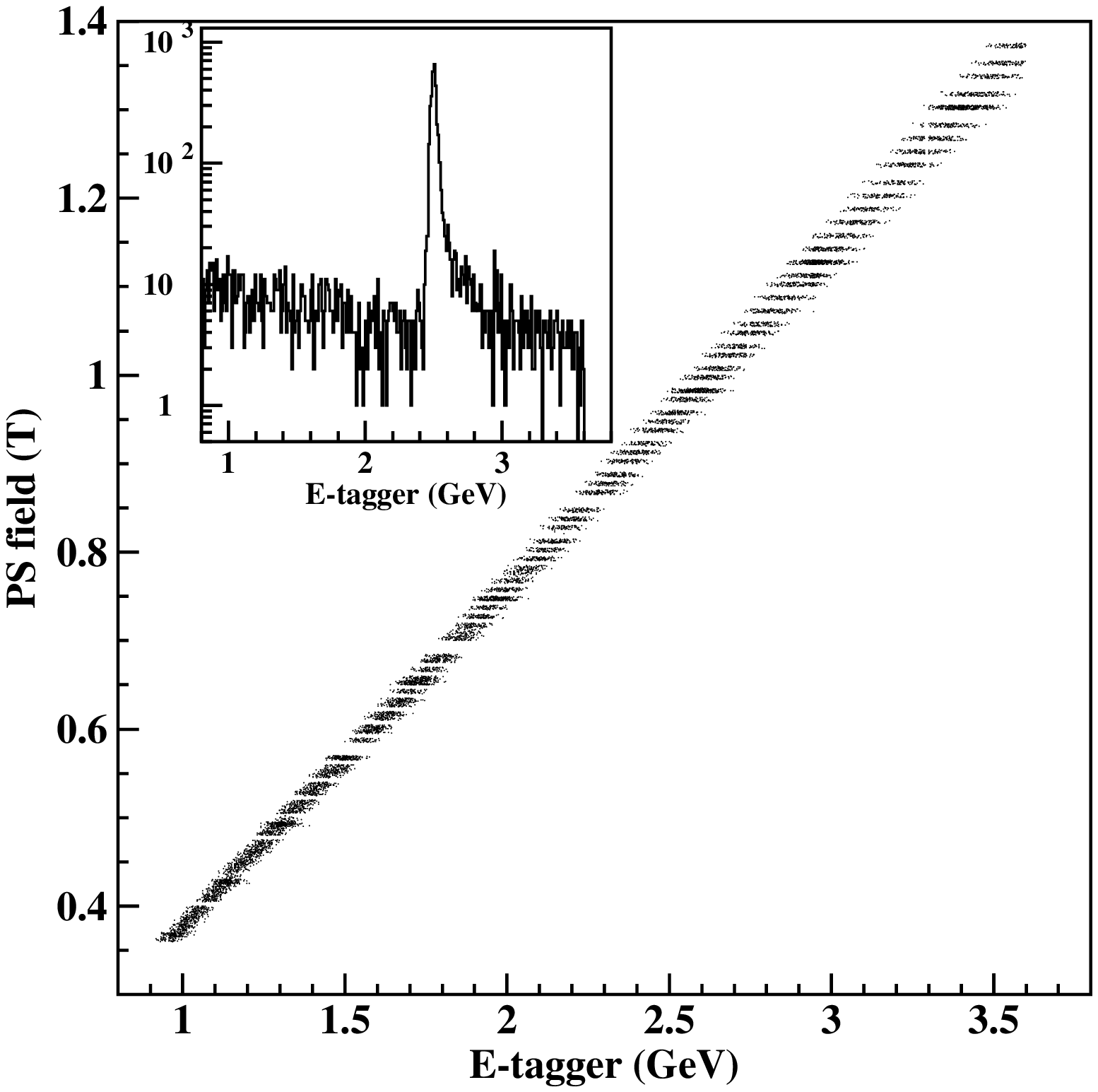}}
\caption{Scatter plot of tagged photon
energy values for each pair spectrometer dipole magnetic field setting. The
inset shows the tagged photon energy distribution at the 
PS dipole field of $0.97$ T.}
\label{fig:be}
\end{figure}

The measurements were conducted at a large number of settings of the 
PS dipole field in the range from $0.36$ to $1.3$ Tesla
which corresponds to PS currents from $543$A to $2278$A. At this range
of the field values, the energy range of $e^+e^-$ covered almost the
full energy range of the tagging system, see Fig. \ref{fig:be}. The PS
magnetic field value was 
measured with a Hall probe positioned at
the center of the magnet. The accuracy of the
device in the range of the measured fields is better than 
$10^{-3}$. The entire data taking process was automated. At each field
setting, data were acquired for $15$
minutes of real beam time with a beam current $>5$ nA. A total of
$180$ points at different field values were measured.

For the determination of the absolute energy scale, data were taken
at PS field values from $1.35$ to $1.9$ Tesla, corresponding to
currents from $2073$A 
to $2775$A, without requiring a coincidence with the tagging system.   
At these field values, the sum of energies
of the detected $e^+$ and $e^-$ covered the range slightly below and
above the electron beam energy (the end-point of the bremsstrahlung spectrum).
Measurements of the $e^+e^-$ coincidence rate at a fixed acceptance of 
the detectors allowed us to relate the PS field values
to the electron beam energy. Using this relation, the absolute energy
scale in the pair spectrometer was defined.

\section{Data analysis}
\indent

A coincidence signal from the scintillator counters of the pair
spectrometer was used to form a trigger for the DAQ system. For each
trigger, time information from the tagger $E$- and
$T$-counters and the amplitude of the signals in each channel 
of the micro-strip detectors were recorded. Information
on beam current,  beam position, and magnetic field settings
was inserted into the data stream every 10 seconds during data taking.

In the off-line analysis, events with tagger hits within 15ns
relative to the trigger were used. Hits in the tagger are selected using 
a tight timing coincidence between $E$-counters and the
corresponding $T$-counter. The inset in Fig. \ref{fig:be} shows
the tagged
photon energy distribution at a single setting of the PS dipole field
(0.97 T) for selected tagger hits. The accidental background in the
tagger was of the order of $1\%$. 

\begin{figure}[htb]
\vspace{80mm} 
{\includegraphics{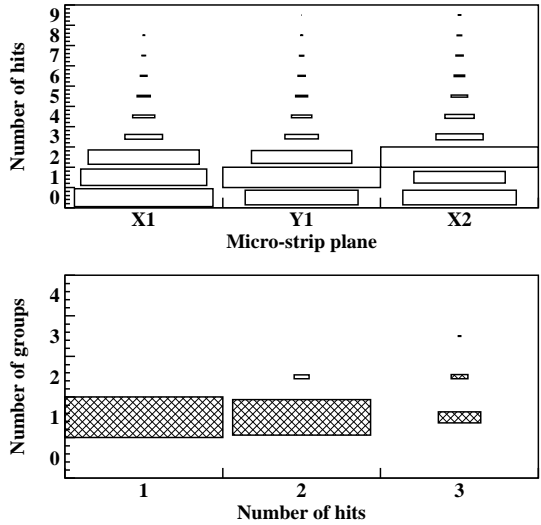}}
\caption{Top panel: the distribution of the number of hits for the $X1$, $Y1$, and $X2$
planes of the micro-strip detector. Bottom panel: the distribution of
the number of adjacent hits
(``groups'') as a function of the number of hits for $1$, $2$, and $3$ hits
cases of $X1$ plane. Shaded boxes are the combinations that were
allowed for the analysis. The size of the boxes 
corresponds to the number of events.}
\label{fig:hits}
\end{figure}

For the determination of the intersection point of $e^+e^-$ trajectories
with the plane of micro-strip detectors, only
channels with ADC values that were above the pedestal by more than five
standard deviations were selected. 
Valid hits in both ``X'' planes and in one  ``Y'' plane were required. 
The distributions of the number of hits in these planes are shown in
Fig. \ref{fig:hits}. A small fraction of events with more than $3$ hits
in $X1$, $Y1$, or $X2$ planes was rejected.
For the final analysis, events with $2$ or $3$ hits in a
given plane were accepted if $2$ of these hits were adjacent (the requirement of
adjacent hits rejected only few events). In the
bottom panel of Fig. \ref{fig:hits}, 
the number of adjacent hits vs. the number of hits on the $X1$ plane is
plotted. The distributions in the $Y1$ and the $X2$ planes were
similar. The hatched boxes on the graph 
correspond to the criteria for the event selection. For adjacent hits, the position
on the plane is calculated as a weighted average using the ADC values.

To reduce uncertainties due to the PS angular acceptance, the
$e^+e^-$ scattering plane was 
limited to $\pm 0.25$ cm around the detector mid-plane.

\clearpage

\section{Tagger energy corrections}

\indent

The derivation of corrections to the tagger energy is performed in two
steps. First, the mean values of the ratio of the photon energy,
measured in the PS,
to the central value of the $E$-bins were calculated. Then, these mean
values were scaled by the ratio of the electron beam energy to the
end-point energy measured in the PS.  

The photon energy is defined as a sum of the $e^+e^-$ energies
reconstructed in the PS. The energies of the electron and the positron
were reconstructed in a simplified model for charged particle 
propagation through the magnetic field. The ratios were
calculated on an event-by-event basis for each E-bin that was within the
acceptance range of the PS at a given field setting. 

The scale factor for the absolute energy determination in PS was derived
from studies of the $e^+e^-$ coincidence rate as a function of the energy
reconstructed in PS.

\subsection{Model for photon energy reconstruction in PS}

\indent

In Fig. \ref{fig:model}, the particle trajectories passing through the
magnetic field of the PS dipole magnet are shown by the solid-lines. For
each event, the transverse displacements of the trajectories 
from the beam centerline ($d$) for the $e^+$ and the $e^-$  were
measured in the micro-strip detector plane. The momenta of leptons
($P$) were calculated in a model that assumes a uniform field 
distribution between the two points along the trajectory 
of the particle. 
In this approximation the momentum can be expressed as a
linear function of the magnetic field strength, $B$, and the radius of
curvature, $R$:
\begin{eqnarray}
P&=& 0.2997925\cdot R\cdot B_0
\label{eq:pe}
\end{eqnarray}

\noindent
where $P$ is in GeV/c, $B$ in Tesla, and $R$ in meters.
The radius of curvature is defined as: 
\begin{eqnarray}
R=L_{eff}\cdot \sqrt{\left ({l_p \over {d_e}}\right )^2+1}
\label{eq:r}
\end{eqnarray}
\noindent
where $L_{eff}$ is the effective field length, $l_p$ is the
distance from the magnet center to the detector 
plane, and $d_{e}$ is the transverse displacement of the
trajectory at the detector plane. 

\begin{figure}[ht]
\vspace{80mm} 
{\includegraphics{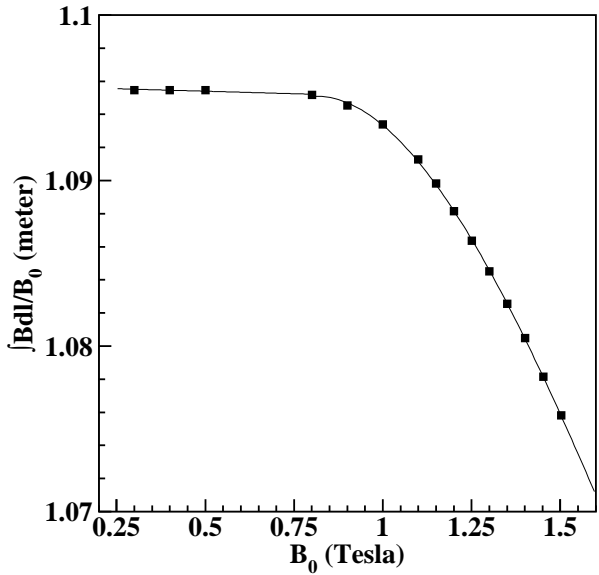}}
\caption{Dependence of $L_{eff}$ on $B_0$, data points. The
  solid line is the fitted function used in the analysis.}
\label{fig:effl}
\end{figure}

The effective field length is defined as:
\begin{eqnarray}
L_{eff}={\int Bdl\over {B_0}}
\label{eq:effl}
\end{eqnarray}
\noindent
were the integral is measured along the trajectory and $B_0$ is the field
value in the center of the magnet. 
The ratio was calculated using the simulation of 
trajectories by the Runge-Kutta-Nystroem method and the field
map generated using a TOSCA \cite{tosca} model for the magnet. 
The generated field map was used in the simulation due to the limited
number of measured 
points for the magnetic field in the acceptance region of the setup. 
Figure \ref{fig:effl} shows the dependence of $L_{eff}$ on
the magnetic field for the
central trajectory inside the acceptance region of the
detector. The drop of $L_{eff}$ at high $B_0$ is due to saturation
effects. It was parametrized  
using a third order polynomial function for the range of field values
$>0.8$ T and a linear function for the range $<0.8$T. The parameterization
was then used to calculate the particle momentum. The effects of the
finite detector sizes, detector geometry, and the size of the beam on
$L_{eff}$ were convoluted into a correction function,
$G(d_{e^+},d_{e^-})$, as explained below. 
For data analysis, another correction function, $F(B_0)$, was introduced to
account for the difference between the generated and the real field
distributions. 
 
The photon energy, $E_\gamma^c$, was defined as a sum of the $e^+$ and
the $e^-$ energies with correction factors  $G(d_{e^+},d_{e^-})$
and $F(B_0)$.

\begin{eqnarray}
E_\gamma&=& (E_{e^+}+E_{e^-})\cdot G(d_{e^+},d_{e^-})\cdot F(B_0)
\label{eq:eg}
\end{eqnarray}

As shown below, using
the simulated field, the accuracy of the approximation of
Eq.(\ref{eq:eg}) is better than the required accuracy for these
measurements ($< 10^{-3}$).

\subsection{Correction for the detector geometry}

\indent

\begin{figure}[ht]
\vspace{80mm} 
{\includegraphics{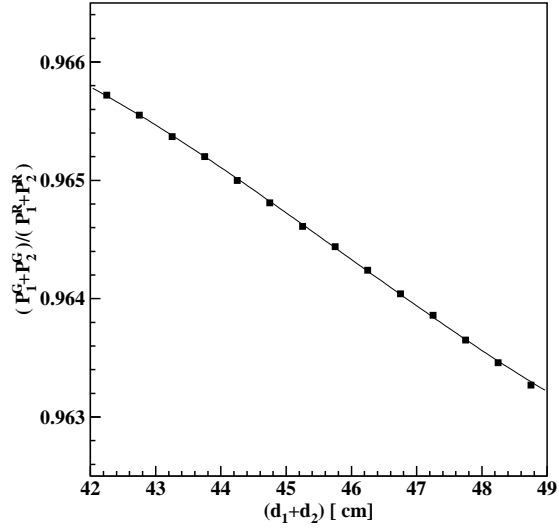}}
\caption{Dependence of the ratio of the total momenta of the simulated 
pair to the sum of the reconstructed momenta, as defined in
Eq.(\ref{eq:pe}), on ($d_1 + d_2$). The central field value was $B_0=0.3$T}
\label{fig:rgx}
\end{figure}

In order to determine the function $G(d_{e^+},d_{e^-})$, pairs of
opposite sign trajectories, originating from the same point at the
pair converter $T$,
were generated for several PS field values in a large momentum
space. The simulations covered the full energy range of
measurements. The starting points of the 
trajectories were distributed in the transverse direction according to the
photon beam profile, using a Gaussian with $\sigma\simeq 1$ mm. As a
correction function 
$G(d_{e^+},d_{e^-})$, the ratio of the sum of generated momenta to the
sum of reconstructed momenta using Eq.(\ref{eq:pe}) was defined. In
Fig. \ref{fig:rgx} the dependence of this ratio on the distance
between $e^+$ and $e^-$ for the central field value $B_0=0.3$T is
presented. An almost linear 
dependence was observed with very small variations ($<0.3\%$) over a
large range of distances. 
The negative slope of the dependence reflects the fact that
$L_{eff}$ was calculated for the central trajectory that corresponds to
$d_{e^+}+d_{e^-}=45$ cm. For tracks with $d_{e^+}+d_{e^-}<45$ cm
($d_{e^+}+d_{e^-}>45$ cm), $\int Bdl$, and therefore the effective
field length, is smaller (bigger) than for the central
trajectory.
The overall scale of the ratio, $<1$, is due to an asymmetric distribution of the
field between tracks start and end points, the MS plane is located farther
from the magnet center than the pair converter.
It should be noted that in the detector geometry of the
experiment, $d_{e^+}+d_{e^-}$ spans the range 
from $43$ to $47$ cm. A third-degree polynomial function was
used to fit the dependence in Fig. \ref{fig:rgx}.

\begin{figure}[ht]
\vspace{80mm} 
{\includegraphics{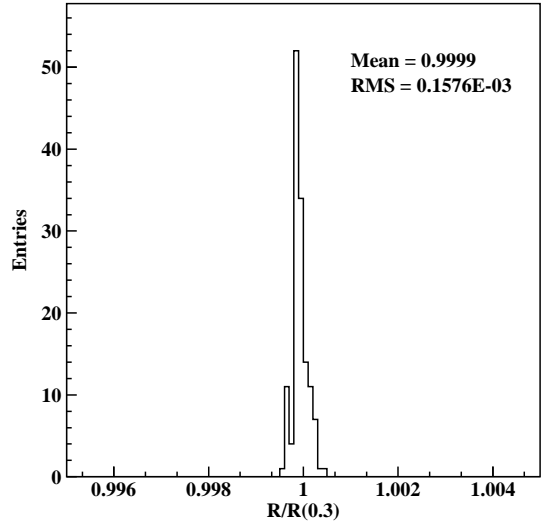}}
\caption{Distribution of the ratio of generated and reconstructed momenta for
several central field values, R, normalized to the ratio at $B_0=0.3$
T, $R(0.3)$, for $15$ points in ($d_1 + d_2$) at each field setting.}
\label{fig:rr03}
\end{figure}

Similar dependences were obtained for several other field
settings. In Fig. \ref{fig:rr03}, the distribution of the ratio
for several field 
values (up to $1.5$ T), divided by the dependence obtained at $0.3$ T,
is presented. The distribution is centered at unity with an RMS value
of $1.6\cdot 10^{-4}$, indicating that the function
$G(d_{e^+}+d_{e^-})$ found at $0.3$ T  
describes the position dependence at all fields very
well. 
%The accuracy of the approximation used in Eq.(\ref{eq:pe}) is much better than $0.1\%$. 

\subsection{Correction for using a simulated field}

\indent
 
\begin{figure}[ht]
\vspace{80mm} 
{\includegraphics{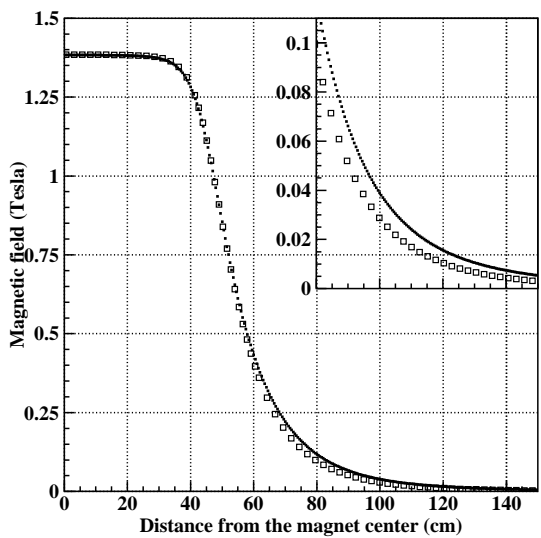}}
\caption{Comparison of the measured and TOSCA-calculated fields.}
\label{fig:tosca}
\end{figure}

\begin{figure}[ht]
\vspace{90mm} 
{\includegraphics{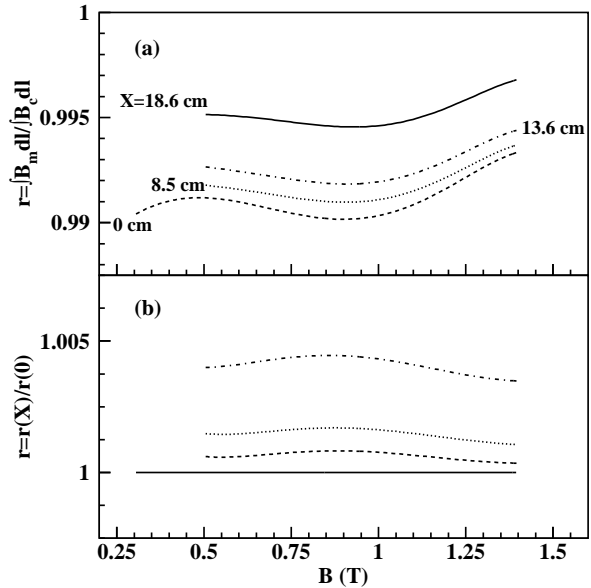}}
\caption{Comparison of $\int Bdl$ values along the $Z$ axis, calculated using the measured
  and the TOSCA-calculated field distributions.
  (a) - the ratio of integrals at different distances from the magnet
  centerline: the dashed line at $X=0$, the dotted line at $X=8.5$
  cm, the dashed-dotted line at $X=13.6$ cm, and the solid line at
  $X=18.6$ cm. In (b) - the same ratios normalized to the ratio at $X=0$.}
\label{fig:rint}
\end{figure}

The effective field length was calculated based on the TOSCA-generated
field map \cite{tosca}. In Fig. \ref{fig:tosca}, a comparison of the
generated (dots) and the measured (open squares) field distributions
is presented. 
There is a small difference in the fringe field and, therefore,
another correction factor was 
introduced to account for this difference. To derive this correction factor, 
the $\int Bdl$ values along the $Z$-axis for the measured and the
TOSCA-calculated field distributions were studied. 
Since the number of points where the B-field was measured was too small
to define the integral along the real trajectories, we compared
integrals along the Z-axis. 
The ratio of these integrals was studied for
four different transverse positions, $X=0$ cm, 
$X=8.5$ cm, $X=13.6$ cm, and $X=18.7$ cm. In Fig. \ref{fig:rint}.a, the
dependence of the ratio on the magnetic field value is shown.
For the $X=0$ point, the $Z$-dependence of the field was measured at 
eight settings and, therefore, the ratio was defined for eight field
settings. At $X=8.5$ cm, $X=13.6$ cm and $X=18.7$ cm, 
measurements are available for only five settings of the PS field. 
The absolute scale of the ratio depends on $X$, but the shape
is similar for all $X$ values. Since the real trajectories
cover a larger range of $X$, and the absolute energy scale will
be defined by the end-point measurements (see below), the absolute
scale of the ratio is
not important. For the analysis, the shape of the fitted dependence,
$F(B)$, at $X=0$ was used. The uncertainty in the
determination of the $F(B)$ is one of the largest contributions to the uncertainty in
the final corrections. As shown in Fig. \ref{fig:rint}.b, the
estimated uncertainty of $F(B)$, 
based on the variations of $r(X)/r(0)$ for single $X$, is $\pm 0.05$\%.

\subsection{Determination of the energy ratios}

\indent

\begin{figure}[ht]
\vspace{80mm} 
{\includegraphics{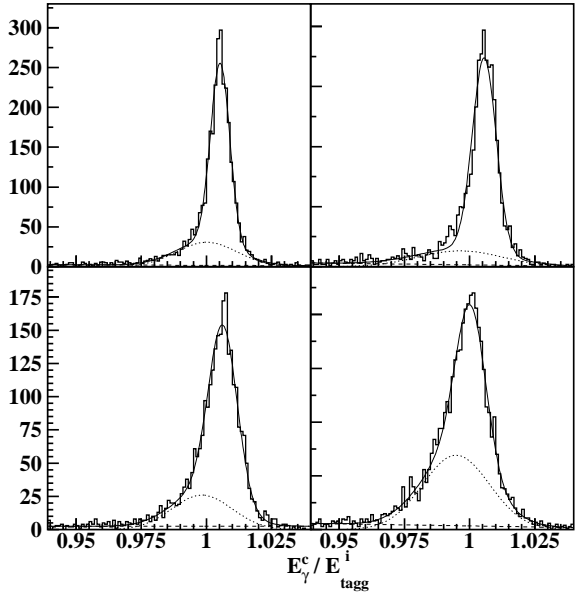}}
\caption{Fit to the $E_\gamma/E^i_{tag}$ distributions for $E$-counters \#{\tt 76},
  {\tt 149}, {\tt 391}, and {\tt 576}. As a fit function a sum of two
  Gaussian and a linear function is used. The mean of the narrow
Gaussian, that describes the main peak, was used to determine the correction
factor to the tagger energy, $C_i$.}
\label{fig:fite}
\end{figure}

The distributions of the ratio
$E_\gamma/E_{tag}^i$ are shown in Fig. \ref{fig:fite} for $E$-bins
$i=76$, $149$, $391$, and $576$. Similar distributions have been analyzed for all
$E$-bins. The distributions are fitted to a sum
of two Gaussians and a linear function. In the figure, fit results are
shown with lines. The mean value, $C_i$, of the narrow Gaussian, that describes the main
peak, was used to determine the correction factor to the tagger energy.

\section{End-point measurements}

\indent

For a given detector acceptance, an increase of the magnetic
field value selects $e^+e^-$ 
pairs from higher energy photons, and at some point reaches the
end-point energy, $E_0$. Measurements of the $e^+e^-$ coincidence
rate as a function of the field value at fixed geometry allow to
determine the relation between field value and the end-point
energy of the photon beam. For these measurements the maximum energy
of photons, or the energy of the electron beam, was $E_0=3.776$
GeV. This value was determined based on independent energy measurements
performed in Hall A and in the accelerator. The difference between
these two measurements was $\sim 3\times 10^{-4}$ and was taken as
the accuracy of the electron beam energy determination. 

\begin{figure}[ht]
\vspace{80mm} 
{\includegraphics{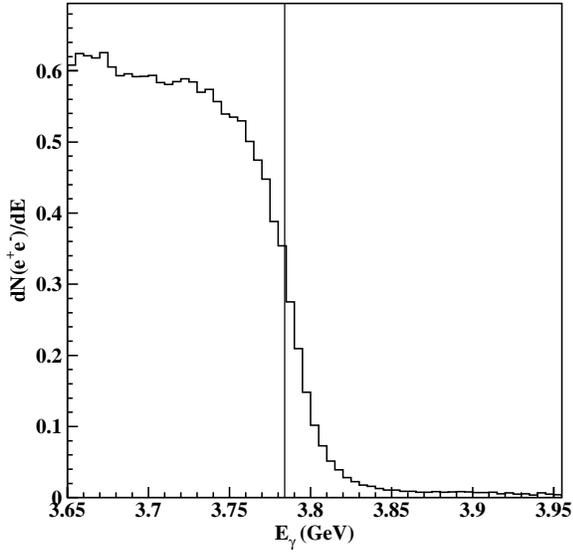}}
\caption{The $e^+e^-$ coincidence rate as a function of energy
  calculated using Eq.(\ref{eq:eg}). The vertical line passes through
  the mid point of the falling edge, and is at $E=3.784$ GeV.}
\label{fig:endp}
\end{figure}

The $e^+e^-$ coincidence rate was studied for several different
detector geometries (different regions of $X$ planes). In Figure
\ref{fig:endp}, the $e^+e^-$ coincidence 
rate is plotted as a function of the photon energy, as defined in
Eq.(\ref{eq:eg}). Data obtained at the PS dipole field values from
$1.35$T to $1.9$T were combined. The shape of the end-point falloff
is defined by the detector resolution. Radiative effects do not
play a significant role at these energies \cite{Schultz}. The deviations
from the lowest order bremsstrahlung cross section at photon energies
$\sim 0.999E_0$ is estimated to be $<10\%$. The energy value
corresponding to the mid point of the falling edge, $E_B=3.784$ GeV,
was taken as the end-point energy. The accuracy of this
approximation was estimated  to be of the order of $5\times 10^{-4}$,
using the results from studies with 
different detector acceptances and taking into account the $10\%$ effect from
radiative corrections. 
A scale factor $\eta=E_0/E_B=0.9985$ was found as
a correction to the energy in Eq.(\ref{eq:eg}), for the PS settings
that were used to derive the correction factors $C_i$.

\section{Final corrections}

\indent

\begin{figure}[ht]
\vspace{80mm} 
{\includegraphics{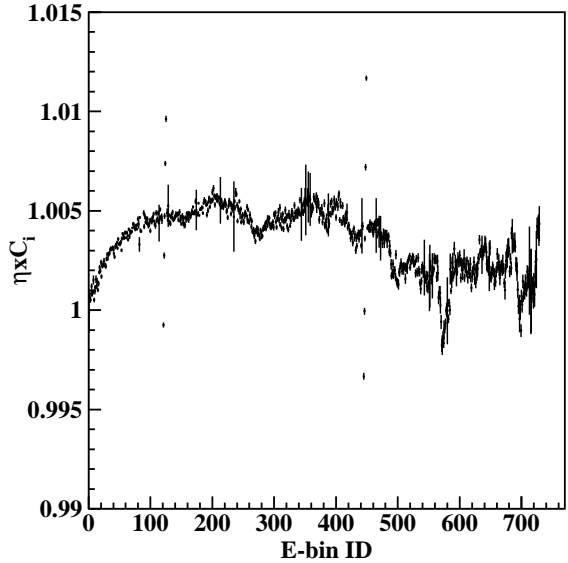}}
\caption{The final tagger energy corrections for each $E$-bin. Smaller
E-bin numbers correspond to higher photon energies.}
\label{fig:corr}
\end{figure}

The final tagger energy corrections were computed using the
$C_i$ constants for each $E$-bin and the energy scale correction
factor $\eta$.
In Figure \ref{fig:corr}, the final corrections are plotted as
a function of the tagger energy bin. A few outlying points around the
E-bins 120 and 440 are due to mis-cablings (which were found in this
measurement and were fixed at a later time).

The estimated error in the determination of the $C_i$ constants is
$0.12\%$. It is defined by the accuracy of the approximation used in
Eq.(\ref{eq:pe}), $1.6\times 10^{-4}$, by the determination of
$F(B_0)$, $5\times 10^{-4}$, by the error in the fit to the
$E_\gamma/E^i_{tag}$ distributions, $\sim 2\times 10^{-4}$, and by the
accuracy of the PS field measurement, $<10^{-3}$. The fit error has a 
small energy dependence. It is small for high energy bins, $1.7\times
10^{-4}$, and larger for th elow energy bins,$3\times 10^{-4}$.
The error in the calculation of $\eta$
arises from the determination of the end-point energy, $5\times
10^{-4}$, and the knowledge of the electron beam energy, $3\times
10^{-4}$. 
For the final correction constants the estimated total uncertainty
is $0.13\%$.

\begin{figure}[ht]
\vspace{90mm} 
{\includegraphics{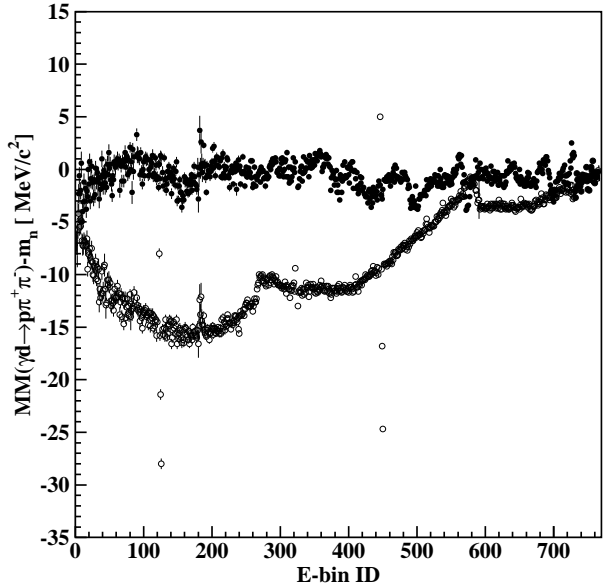}}
\caption{The difference between the missing mass of the
($p\pi^+\pi^-$) system in the reaction $\gamma
d\to p\pi^+\pi^-(n)$ and the nominal neutron mass as a function of
the tagger $E$-bin. Open symbols correspond to the missing mass calculation
without and the filled symbols with the new tagger energy corrections.}
\label{fig:gdppipi}
\end{figure}

In summary, we performed an energy calibration of the Hall B bremsstrahlung
photon tagging system at Jefferson Lab. The calibration was done using the
Hall B pair spectrometer, instrumented with micro-strip detectors for
high precision position measurements. The calibration results were
checked using exclusive photoproduction reactions detected in
CLAS. A result of such an analysis is
presented in Fig. \ref{fig:gdppipi}. The exclusive reaction $\gamma
d\to p\pi^+\pi^-(n)$ was 
studied where the final state proton and two pions were detected in
CLAS, and the neutron was reconstructed from the missing momentum and
missing energy analysis.  
In the figure, the difference of the missing mass of ($p\pi^+\pi^-$)
and the nominal mass of the neutron from the Particle Data Group (PDG)
\cite{pdg} is plotted as a function of tagger E-bin. The open
symbols correspond to the missing mass obtained
without tagger energy 
corrections, and the filled symbols correspond to results when the
tagger energy corrections 
were applied. The variation of the neutron missing mass 
without the corrections are up to $15$ MeV, while with
corrections these variations are within $\pm 3$
MeV. 

%{\bf
The analysis presented in \cite{s1} was repeated with the new
calibration constants and showed similar improvements in the energy
determination. The data in \cite{s1} were from a 1999 CLAS run, while
the data used in this analysis and the data presented in
Fig. \ref{fig:gdppipi} were taken in 2004. Comparison of
these two sets of data shows that the geometry of the tagger focal
plane did not change during this five year period and that the sagging and
misalignments were introduced at the time of the tagger construction.
%}

\vspace{0.5cm}

We would like to acknowledge the outstanding efforts of the staff of
the Accelerator and Physics Divisions at Jefferson Lab who made this
measurements possible. We particularly wish to thank the Hall B technical
staff for exceptional work in the installation of the
experiment. Acknowledgments for the support go also to the National
Science Foundation, MRI 
grant PHY-0079840, grants PHY-0457246, PHY-0099487, and
PHY-0072391. This work authored by The Southeastern Universities 
Research 
Association, Inc. under U.S. DOE Contract No. DE-AC05-84150. The
U.S. Government retains a non-exclusive, paid-up, irrevocable,
world-wide license to publish or reproduce this manuscript for
U.S. Government purposes.


\begin{thebibliography} {999}

\bibitem{tagger}
D.I. Sober et al., Nucl. Instr. Meth. A {\bf 440}, 263 (2000).

\bibitem{clas}
B.A. Mecking et al., Nucl. Instr. Meth. A {\bf 503}, 513 (2003).  

\bibitem{s1}
S. Stepanyan, CLAS-ANALYSIS 2003-105.

\bibitem{taggeom}
D.I. Sober, H. Crannell and F.J. Klein, CLAS-NOTE 2004-019.

\bibitem{primex} The Hall B pair spectrometer was developed and
constructed by the PrimEx collaboration at JLab, http://www.jlab.org/primex/.

\bibitem{tosca}
OPERA-3D User Guide, Vector Fields Limited, 24 Bankside, Kidlington,
Oxford OX5 1JE, England. 

\bibitem{Schultz} H.D. Schultz and G. Lutz, Phys.Rev. {\bf 167}, 1280
(1968). 

\bibitem{pdg} Particle Data Group, Phys. Lett. B 521 (2004).


\end{thebibliography}
\end{document}